\documentclass{article}
\usepackage{newpasp}
\usepackage{graphicx}
\begin{document}

\title{Formation  and  Evolution  of  a  0.242 $M_{\odot}$ Helium
White Dwarf in Presence of Element Diffusion}

\author{    Althaus    L.    G.\altaffilmark{1},   Serenelli   A.
M.\altaffilmark{2} and Benvenuto O. G.\altaffilmark{3}}

\affil{Facultad  de  Ciencias  Astron\'omicas  y Geof\'{\i}sicas,
Paseo del Bosque S/N (1900) La Plata Argentina}

\altaffiltext{1}{Member   of   the   Carrera   del   Investigador
Cient\'{\i}fico    y    Tecnol\'ogico,    CONICET,     Argentina}
\altaffiltext{2}{Fellow    of    the    Consejo    Nacional    de
Investigaciones   Cient\'{\i}ficos    y   T\'ecnicas,    CONICET,
Argentina}   \altaffiltext{3}{Member    of   the    Carrera   del
Investigador  Cient\'{\i}fico,   Comisi\'on  de   Investigaciones
Cient\'{\i}ficas  de   la  Provincia   de  Buenos   Aires  (CIC),
Argentina \\emails: althaus,serenell,obenvenuto@fcaglp.fcaglp.unlp.edu.ar}

\abstract{A 0.242 $M_{\sun}$ object that finally becomes a helium
white dwarf is  evolved from Roche  lobe detachment down  to very
low luminosities.  In doing  so, we  employ our  stellar code  to
which we have  added a set  of routines that  compute the effects
due  to   gravitational  settling,   and  chemical   and  thermal
diffusion. Initial model is constructed by abstracting mass  to
a 1 $M_{\sun}$ red giant branch  model up to the moment at  which
the  model  begins  to  evolve  bluewards.  We  find that element
diffusion introduces noticeable changes in the internal structure
of  the   star.  In   particular,  models   undergo  {\it  three}
thermonuclear  flashes  instead  of  one  flash as we found with the
standard treatment.  This fact  has a  large impact  on the total
mass fraction of hydrogen left in the star at entering the  final
cooling track. As  a result, at  late stages of  evolution models
with diffusion are characterized by a much smaller nuclear energy
release, and  they evolve  significantly faster  compared to 
those found with  the standard treatment.

We  find  that  models  in  which diffusion is considered predict
evolutionary  ages   for  the   white  dwarf   companion  to  the
millisecond  pulsar  PSR  B1855+09  in  good  agreement  with the
spin-down age of the pulsar. \endabstract

\section{INTRODUCTION} \label{sec_intro}

Low mass, helium white dwarfs  (WDs) have drawn the attention  of
many researchers who have devoted a large effort to their  study.
Currently, it  is well  established that  helium WDs  should have
been formed  in a  binary system  because an  isolated star would
reach such a configuration in  a time-scale much longer than  the
present age of our Universe. Examples of recent studies on helium
WD evolution are Althaus \& Benvenuto (1997; 2000), Benvenuto  \&
Althaus (1998), Hansen  \& Phinney (1998),  Driebe et al.  (1998)
and Sarna, Antipova \& Ergma (1999).

A crucial  issue in  the study  of helium  WDs is  the amount  of
hydrogen  left  atop  the  WD  after  the  end  of  mass transfer
episodes.  Sarna  et  al.  (1999)  have  presented  detailed
calculations of the pre -  WD evolution during the mass  transfer
stage  in  a  close   binary  system.  They  found   that,  after
detachment,  helium  WDs  are  surrounded  by  massive   hydrogen
envelopes of  0.01 -  0.06 $  M_{\sun}$ with  a surface  hydrogen
abundance  by  mass  of  $X_{H}=$  0.35  -  0.5. Massive hydrogen
envelopes have  also been found  by  Driebe et  al. (1998), who
have mimicked the binary evolution by abstracting mass from a  $1
M_{\sun}$ model at an adequate rate.

Binary systems composed by a  helium WD and a millisecond  pulsar
offer us  a  good  opportunity  to  test  the  predictions  of   
evolutionary calculations. Millisecond pulsars are thought to  be
recycled during the mass transfer stage (Bhattacharya \& van  den
Heuvel 1991). When mass transfer ends, the pulsar begins to  spin
down. If we set zero ages at the end of such mass transfer,  then
the ages of the WD and  the pulsar component should be the  same.
If the pulsar radiates as a dipole and its current period ($P$) is much
longer than the initial one, then the spin-down of the pulsar  is
given by $t_{PSR}= P/2\dot{P}$, where $\dot{P}$ is the rate of change
of $P$.

Recently, van Kerkwijk et al. (2000) have optically detected  the
low-mass component of the PSR B1855+09 system and determined  its
effective temperature to  be $4800 \pm  800$ K. In  addition, the
mass of the WD is accurately  known thanks to the measure of  the
Shapiro  delay   of  the   pulsar  signal   and  results   to  be
$0.258^{+0.028}_{-0.016} M_{\sun}$. The age  derived for the  WD component
results to be twice as long as the pulsar spin-down age  ($\approx
5  Gyr$)  when  using the models without diffusion of Driebe et al. (1998).

Helium WD evolution  in presence of  element diffusion has  been
considered by Iben \&  Tutukov (1986), who computed  the evolution
of a $0.296 M_{\sun}$  object throughout the entire  evolutionary
stages  and  found  two  thermonuclear  flashes. Remarkably, they
found that after flashes the  amount of hydrogen in the  envelope
of the model was unable to support an appreciable amount of  hydrogen
burning reactions, enforcing a  fast evolution for the  resulting
WD star.

It is the  purpose of this  work to study  the impact of  element
diffusion on the  structure and evolution  of a $0.242  M_{\sun}$
helium WD star. This mass  value is  representative of  the
companion to  PSR B1855+09.  In doing  so, two  sequences are
computed: one  with and  the other  without diffusion.  Thus, the
resulting differences should be due  to the allowance for such  a
process.

\section{EVOLUTIONARY RESULTS} \label{sec_resu}

For the present  work, we shall  employ the same  stellar code we
used in our previous works on WD evolution (Althaus \&  Benvenuto
1997; Benvenuto \& Althaus 1998). For the computations  presented
in this paper, we have included gravitational settling,  chemical
and thermal diffusion, as described by Althaus \& Benvenuto
(2000), and we refer the reader to that paper for further details.

In order to have a  realistic starting model, we have  abstracted
mass to a $1  M_{\sun}$ giant star up  to the moment at  which it
begins to evolve  bluewards. In this way we have  simulated
the mass transfer episode  during Roche lobe overflow.  From then
on,  the  evolution  of  the  model  has been followed assuming a
constant    value    for    the    stellar    mass    down     to
$\log{L/L_{\sun}}=-5$.  The  results  we  have calculated without
diffusion  are  in  very  good  agreement with those presented by
Driebe et al. (1998), thus in comparing the results we have found
considering and neglecting diffusion we are indeed also comparing
with the results of Driebe et al.

\begin{figure}[h] 
\vskip -1cm
\hskip -4cm
\includegraphics[height=8in,angle=-90]{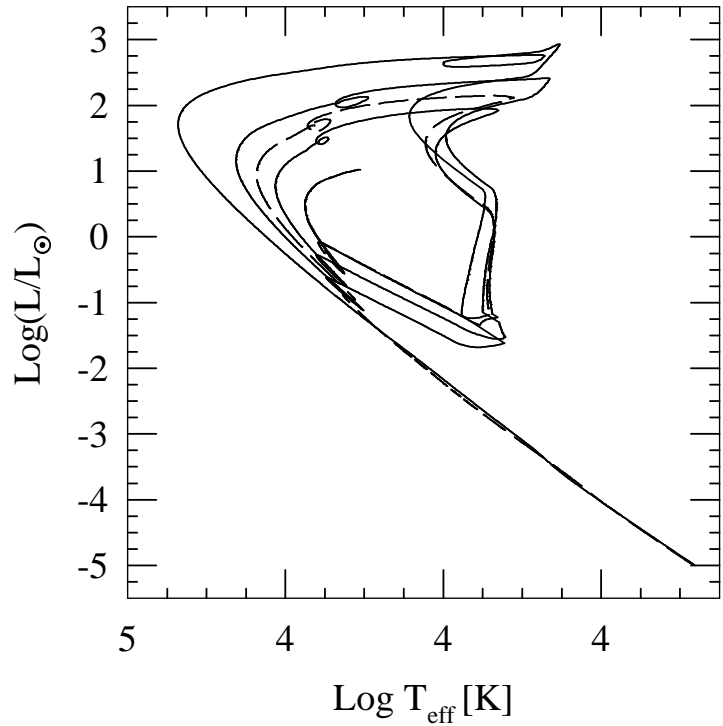}
\vskip -1cm
\caption{Evolutionary tracks for a
0.242  $M_{\sun}$  object  which  finally  reaches  a  helium  WD
structure  considering  and  neglecting  element diffusion. Solid
line corresponds to models  with diffusion, whereas short  dashed
line  depicts  standard  models  without  diffusion.  Notice that
models with diffusion  undergo three loop  excursions to the  red
giant part of  the HR diagram,  whereas models without  diffusion
experience only one of such episodes.} 
\end{figure}

The evolutionary tracks for the models with and without diffusion
are shown in  Fig. 1. While  the standard treatment  predicts the
occurrence  of  one  loop,  models  with  diffusion undergo three
loops. Each of these loops are due to unstable nuclear burning at
the  bottom  of  the  hydrogen-rich envelope, i.e., thermonuclear
flashes.  These  flashes  are  critical  events  that, to a large
extent, determine  the subsequent  evolution of  the object, even
during  the  final  WD  cooling  phase.  The  occurrence of three
flashes in models  with diffusion is  in sharp contrast  with the
predictions of the standard treatment, and clearly indicate  that
element diffusion is by no  means a negligible phenomenon in  the
context of such stars. After flashes there are somewhat long time
periods  for  which  both nuclear and photon  luminosities  are  
high. These periods
correspond to evolutionary stages  during which the total  amount
of hydrogen in the star ($M_{H}/M_{*}$) is noticeably reduced, as
it is apparent from Fig. 2.

\begin{figure}[h] 
\vskip -3cm
\centering
\includegraphics[height=5in]{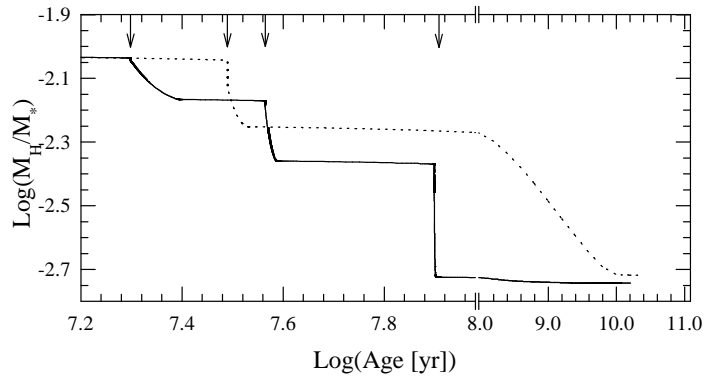}
\vskip -4cm
\caption{The total mass  fraction
of hydrogen as a function of time for the 0.242 $M_{\sun}$ helium
WD model considering and neglecting element diffusion. Solid line denotes models
with diffusion and dotted line corresponds to models without diffusion.
Ages  corresponding  to the
occurrence of  thermonuclear flashes  are indicated  by arrows. A
noticeable hydrogen depletion is found immediately after flashes.
Notice  that  models  without  diffusion  enter the final cooling
branch ($t > 1 Gyr$) with  a much higher hydrogen mass, which  is
burnt  during  this  epoch.  In  models  with  diffusion, nuclear
reactions also occur, but at  a much lower total rate.  This fact
has a large impact on the final age of the star.} \end{figure}

An appreciable  hydrogen envelope  reduction is  also found after
the only flash  in the sequence  of models without  diffusion but
the key difference between the two sets of models is related with
the occurrence  of nuclear  burning during  the final  WD cooling
phase. In models with diffusion and after flashes, the total mass
fraction  of  hydrogen  left  is  $M_{H}/M_{*} \approx 1.8 \times
10^{-3}$. However, in the case of models without diffusion,  when
the star enters  the cooling track,  the amount of  hydrogen that
remains  is  about  three  times  as  large  as  in the case with
diffusion. As a direct
consequence of such a large difference, models without  diffusion
burn $\approx$
2/3 of its hydrogen content {\it during the cooling  track}.
In models with diffusion, nuclear reactions also occur, but at  a
much lower total rate. This fact  has, as we shall show below,  a
large impact on the final age of the star.

\begin{figure}[h]
\vskip -2cm
\hskip -1cm
\includegraphics[height=6.5in,angle=-90,viewport=0 20 500 800]{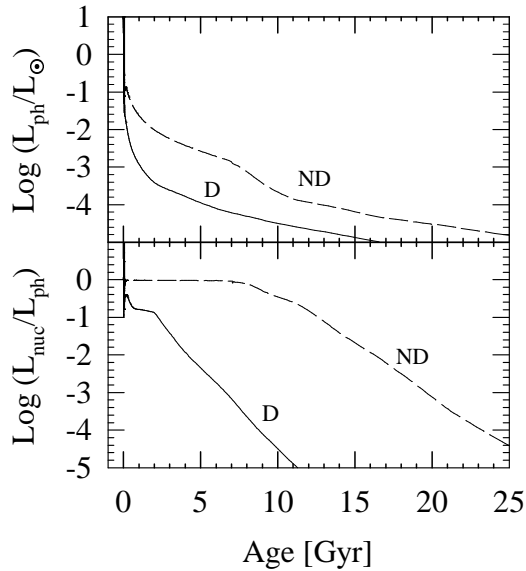}
\vskip -1cm
\centering
\caption{Photon and nuclear luminosity  vs. age
relationship for models considering (solid lines) and  neglecting
diffusion (dashed lines).  Clearly, models with  diffusion evolve
much more rapidly because the  hydrogen envelope left is so  thin
that it is unable to support a appreciable amount of nuclear burning
(lower panel). This is  in contrast with standard  results without
diffusion for which  hydrogen burning is  the main energy  source
even at very low luminosities.} \end{figure}

From the results  presented in Fig.  3 (upper panel)  it is clear
that models with diffusion undergo a much faster cooling compared
with  models  without  diffusion.  In  models  without diffusion,
nuclear reactions release an amount  of energy very close to  the
radiated one. This strongly delays cooling to very long ages. 
On the contrary, diffusion leads to hydrogen envelopes thin enough
so as to prevent nuclear reactions from dominating the energetics
of the star. Thus, the
star  has  to  extract  energy  from  the  relic thermal content,
forcing  a  much  faster  cooling  (in  agreement  with   results
presented in Benvenuto \& Althaus 1998). To put this assertion on
a   more   quantitative   basis,   we   note   that  in  reaching
$\log{L/L_{\sun}}=-4$  models  with  diffusion  require  5.3 Gry,
whilst  models  without  diffusion  need  13  Gyr. In view of the
present results, the controversy found by van Kerkwijk et al.
(2000) about the  large discrepancy  between the ages  of the
pulsar PSR B1855+09 and its WD companion is strongly alleviated.

\section{Discussion and Conclusions} \label{secc:conclu}

From the results  presented in the previous section,  it is clear that
element  diffusion  introduces   important  changes  in  the  internal
structure of  the star.  Our  present detailed results  indicate that,
contrary  to Driebe  et al.  (1998) and  Sch\"onberner et  al.  (2000)
assertions,  nuclear  reactions are  not  an  important ingredient  in
determining the  age of the  star during the final  cooling evolution.
This conclusion is valid at  least for the stellar mass value analyzed
in this paper.  Although the hydrogen mass fraction at  the end of the
computed stages  is very  similar in both  sequences with  and without
diffusion,  there is a  large difference  in the  total time  spent in
reaching such  a low  luminosity in each  case. While  standard models
without diffusion  spend about  30 Gyr in  reaching $\log{L/L_{\sun}}=
-5$, models with diffusion get such a luminosity in about 15 Gyr.

As stated  in the introductory  section, previous models  available in
the  literature (those in  which hydrogen  burning is  dominant during
most of the  cooling stages) lead to a  strong discrepancy between the
ages of PSR  B1855+09 and its helium WD  companion.  These ages should
be, on very general grounds,  the same.  The results presented in this
paper  strongly indicate  that {\it  such  a discrepancy  is a  direct
consequence  of  ignoring  element  diffusion  in  the  stellar  model
calculations}. If we  include it properly, ages naturally  come into a
nice agreement. Consequently, we do not have to invoke any ad-hoc mass
loss as proposed  by Sch\"onberner et al. (2000)  or exotic mechanisms
to  account  for the  observational  situation  of  the binary  system
containing PSR B1855+09.

It is worth  noting that the structure of the  models belonging to the
final cooling branch we have found  in the present paper for the 0.242
$M_{\sun}$ helium  WD strongly resemble  those we have assumed  in our
previous  work on  helium WDs  with hydrogen  envelopes  (Benvenuto \&
Althaus 1998).  In fact, if  we assume $M_{H}/M_{*}= 2 \times 10^{-3}$
the  agreement  with  the  present  results  is  excellent,  not  only
regarding the cooling  evolution but also with respect  to the nuclear
activity.   In  our  opinion,  the present  detailed  results  largely
justify the assumptions we have made in our quoted previous paper.

We  want  to  emphasize  that  our  results  strongly  resemble  those
presented sometime ago  by Iben \& Tutukov (1986)  (IT86) for the case
of a  $0.296 M_{\sun}$  model.  They found  two flashes  and suspected
that the  second one was  due to diffusion,  but because they  did not
compute models without diffusion,  they were not conclusive about this
point. Here we confirm that diffusion is the reason for the occurrence
of  additional flashes.  IT86  also  found that  the  final amount  of
hydrogen left in  the WD is small enough to  ultimately enforce a fast
cooling.  Although  their  model  is  more massive  than  ours,  which
prevents from a quantitative comparison between both set of calculations,
the role of  hydrogen burning at the final cooling  stages is the same
we have found.  There  is nevertheless an important difference between
the work of IT86 and our study.  In fact, IT86 considered that shortly
after the  second flash,  when the model  gets its maximum  radius, it
again overfills its  Roche lobe and losses a tiny  part of its stellar
mass. Obviously, this mass belongs to the hydrogen envelope and such a
loss has a large impact on the final value of $M_{H}/M_{*}$ in the WD.
This is ultimately responsible for  the fast cooling occurring at late
stages of  evolution. The point to  emphasize here is that  we did not
consider such a mass transfer  during our calculations, and some words
about this  point are  in order. If  such a transfer  actually occurs,
$M_{H}/M_{*}$  should  drop  below   the  values  we  presented  above
(eventually  to  zero).  This  would  have  an  impact on  the  global
structure of the  WD. Such a situation may be  considered in the frame
of the  models presented in  Benvenuto \& Althaus (1998).   Apart from
the structural differences we do not expect an important change in the
time-scale  of  cooling.  Notice  that  in  the  models  presented  in
Benvenuto \&  Althaus, the time  spent in reaching a  given luminosity
value is rather independent from  the exact value of $M_{H}/M_{*}$ (if
it is in the range of $1 \times 10^{-3}$ to zero). Thus, the occurence
of such a  late mass transfer should not  modify the conclusions about
the  age of  the final  WD  model and  its better  agreement with  the
observations related to the ages of the PSR B1855+09 system.

\acknowledgments We would like to thank H. Shipman and the LOC of
the Twelfth European Workshop on White Dwarfs for their  generous
support that allowed us to be there.

\references

\reference Althaus L. G., \& Benvenuto O. G., 1997, ApJ, 477, 313

\reference Althaus L. G., \& Benvenuto O. G., 2000, MNRAS, 317, 952

\reference Benvenuto O. G., \&  Althaus L. G., 1998, MNRAS,  293,
177

\reference Bhattacharya  B., \&  van den  Heuvel E.  P. J., 1991,
Phys. Rep., 203, 1

\reference Driebe T., Sch\"onberner  D., Bl\"ocker T., \&  Herwig
F., 1998, A\&A, 339, 123

\reference Hansen B. M. S., \& Phinney, E. S., 1998, MNRAS,  294,
557

\reference Iben I. Jr., \& Tutukov A. V., 1986, ApJ, 311, 742

\reference Sarna  M. J.,  Antipova J.,  \& Ergma  E., 1999, 11th.
European Workshop on White Dwarfs (Solheim J.-E., Meistas E.  G.,
Eds.) ASP Conference Series, 169, 400

\reference Sch\"onberner  D., Driebe  T., \&  Bl\"ocker T., 2000,
A\&A, 356, 929

\reference  van  Kerkwijk  M.  H.,  Bell  J.  F., Kaspi V. M., \&
Kulkarni S. R., 2000, ApJ, 530, L37

\endreferences

\end{document}